\documentclass[pra,twocolumn]{revtex4}
\usepackage{latexsym}
\usepackage{amsmath}
\usepackage{amssymb}
\usepackage{theorem}
\usepackage{graphicx,epsfig}
\usepackage{mathrsfs}
\def\bP{{\mathbf P}}\def\bQ{{\mathbf Q}}\def\bR{{\mathbf R}}
\def\bX{{\mathbf X}}\def\bD{{\mathbf D}}
\def\Proof{\medskip\par\noindent{\bf Proof. }}

\newtheorem{theo}{Theorem}
\newtheorem{lemm}{Lemma}

\newtheorem{definition}{Definition}
\newtheorem{corollary}{Corollary}

\def\<{\langle}\def\>{\rangle}\def\Tr{\operatorname{Tr}}
\def\set#1{{\sf #1}}\def\sE{\set{E}}\def\sH{\set{H}}\def\sK{\set{K}}
\def\Cmplx{\mathbb C}\def\Reals{\mathbb R}\def\Span{\set{Span}}
\def\N#1{\left|\!\left|#1\right|\!\right|}\def\ie{i.~e. }
\def\qed{$\,\blacksquare$\par}
\def\vec#1{{\mathbf #1}}
\begin{document}
\title{Quantum indirect estimation theory and joint estimate of all moments of two incompatible observables}

\author{G. M. D'Ariano} 

\affiliation{QUIT Group,
  Dipartimento di Fisica ``A. Volta'', via Bassi 6, I-27100 Pavia,
  Italy and CNISM.} 

\author{P. Perinotti} 

\affiliation{QUIT Group,
  Dipartimento di Fisica ``A. Volta'', via Bassi 6, I-27100 Pavia,
  Italy and CNISM.} 

\author{M. F. Sacchi}

\affiliation{QUIT Group,
  Dipartimento di Fisica ``A. Volta'', via Bassi 6, I-27100 Pavia,
  Italy and CNISM.} 

\begin{abstract}
  We introduce the quantum indirect estimation theory, which provides a
  general framework to address the problem of which ensemble averages
  can be estimated by means of an available set of measuring
  apparatuses, e.~g. estimate the ensemble average of an observable by
  measuring other observable. A main ingredient in this approach is
  that of informationally complete ({\em infocomplete} in short)
  measurements, which allow to estimate the ensemble average of any
  arbitrary system operator, as for quantum tomography. This naturally
  leads to the more stringent concept of $AB$-informationally complete
  measurements, by which one can estimate jointly all the moments of
  two incompatible observables $A$ and $B$. After analyzing all
  general properties of such measurements, we address the problem of
  their optimality, and we completely solve the case of qubits,
  showing that a $\sigma_x\sigma_y$-infocomplete measurement is less
  noisy than any infocomplete one.  We will also discuss the relation
  between the concept of $AB$-infocompleteness and the notion of joint
  measurement of observables $A$ and $B$.
\end{abstract}
\maketitle 
\section{Introduction}
The aim of any measurement is to retrieve information on the state of a physical system. In
classical mechanics, measuring the location on the phase space provides a complete information on
the system. On the other hand, in quantum mechanics there are infinitely many elementary
measurements---corresponding to different observables---that provide only partial information,
whereas ``complementary'' informations would require mutually exclusive experiments corresponding to
non-commuting observables.

The problem then arises on how to perform a quantum measurement that can be used to infer
information on non compatible observables. The idea is to make a generalized ``unsharp'' measurement
\cite{bus}, described by a so-called POVM (positive-operator valued measure), from which a specific
type of information---such as e. a particular ensemble average of a given operator---is retrieved by
a suitable data-processing of its experimental outcomes.

\par Of special interest are the informationally complete POVMs \cite{prug}---{\em
  infocomplete} POVMs in short---which span the whole operator space, thus allowing the estimation of
arbitrary ensemble averages.  Informationally complete measurements are relevant for foundations of
quantum mechanics as a kind of ``standard'' for a purely probabilistic description \cite{fuco}.
Moreover, the existence of such measurements with minimal number of outcomes is crucial for the
quantum version of the de Finetti theorem \cite{caves}.  The most popular example of informationally
complete measurement is given by the coherent-state POVM for a single mode of the radiation field,
whose probability distribution is the so-called $Q$-function (or Husimi function) \cite{cah}.
Another example, though of completely different kind, is the case of quantum tomography \cite{tomo},
in which one measures an observable randomly selected from an informationally complete set---a
"quorum".

\par Investigations on informationally complete measurements have been extensively carried out. In
the framework of ``phase-space observables'' \cite{Holevo,Davies,Busch,Schroeck,Perinova} the
concept of informational completeness leads to substantial advancement on some relevant conceptual
issues, such as the problem of jointly measuring non-commuting observables, or the problem of the
classical limit of quantum measurements.  A general classification of covariant infocomplete
measurements has been given using group-theoretical techniques \cite{jop}, whereas the
classification of the symmetric ones is still an open problem \cite{appleby}. A thorough comparison
of local with global infocomplete measurements for bipartite quantum systems has been carried out in
Ref.  \cite{locvs}. On the other hand, for any general infocomplete measurement the optimal
data-processing function for estimating the ensemble average of an arbitrary operator has been
derived \cite{optproc} with the help of frame theory \cite{ds,czz}.

In this paper we introduce the quantum indirect estimation theory, which provides the general
framework to address the problem of which ensemble averages can be estimated by means of an
available set of measuring apparatuses. Typically, one has the problem of estimating the ensemble
average of an observable by measuring other observables, or of estimating the expectation of a
POVM---i.~e. a probability distribution---by physically measuring another POVM. Essentially, one can
estimate all expectations of operators that are linear combinations of POVM elements. The indirect
estimation is achieved via a {\em data processing} of measurement outcomes.  The data processing
associates a numerical value to each outcome, depending on the ensemble average to be estimated.
The final goal of the theory is then to optimize the data processing (generally not unique) in order
to maximize statistical efficiency \cite{optproc}. A special case of data-processing is the {\em
  post-processing}, which corresponds to probabilistic Boolean operations and permutations on the
outcomes, with the data-processing function corresponding to a conditional probability. A typical
example of post-processing is the {\em coarse-graining} of a POVM, in which each outcome is indeed a
union of elementary outcomes, e.~.g. in the {\em marginalization} of a bi-variate POVM.

Clearly, a central role in quantum indirect estimation theory is played by infocomplete POVM's, by
which one can estimate the ensemble averages of any arbitrary operator. However, for the estimation
of the ensemble averages $\<A\>$ and $\<B\>$ of two (noncommuting) operators one does not
necessarily need an infocomplete measurement, even in the case when one wants to estimate the full
probability distribution of $A$ and $B$. In the last case one just needs a particular measurement,
that we will introduce in the present paper, and which will be referred to as $AB$-infocomplete
measurement. Indeed, the necessity of estimating complementary observables is the reason why the
POVM which achieves the task is unsharp, and whence it adds noise to the POVM which can estimate a
single observable. Likewise, one can infer that an $AB$-infocomplete POVM which is not infocomplete
should add less noise than an infocomplete one, since the first kind of measurement avoids to
collect redundant information. We will see that this indeed is true in the special case of qubits.
We will also see that generally a joint measurement of observable $A$ and $B$ is not necessarily an
$AB$-infocomplete measurement, whereas, viceversa, an $AB$-infocomplete measurement is an unbiased
joint measurement of $A$ and $B$.

The paper is organized as follows. Sec. II is a long section where we introduce the quantum indirect
estimation theory through the notion of partially informationally complete POVM, where the linear
span of the POVM elements is a proper subspace of the Hilbert-Schmidt operator space. We also
briefly review the theory of frames \cite{ds,czz}, which generalize the concept of (operator) basis,
and show how to characterize and optimize the processing functions of quantum measurements to
estimate the expectation of observables. The notion of data-processing and post-processing are
explained, and the concept of joint measurement of observables is recalled. In Sec. III minimal
$AB$-infocomplete measurements are introduced, as the measurements described by POVMs whose span
coincides with the span of $A$, $B$ and all their independent powers. A useful Lemma that gives
sufficient conditions for minimality of the {\em optimal} $AB$-infocomplete measurement is proved.
The case of qubits is solved in Sec. III.A, when the ensemble of unknown states corresponds to an
isotropic distribution. Sec. IV is devoted to the conclusions.

\section{Indirect estimation theory}

A measurement on a quantum system \cite{Holevo} returns a random result $e$ from a set of possible
outcomes $\sE=\{e:1,\ldots N\}$, with probability distribution $p(e|\rho)$ depending on the state
$\rho$ of the system in a way which is distinctive of the measuring apparatus, according to the Born
rule
\begin{equation}
p(e|\rho)=\Tr[\rho P_e].\label{Born}
\end{equation}
In Eq. (\ref{Born}) $P_e$ denote positive operators on the Hilbert space $\sH$ of the system,
representing our knowledge of the measuring apparatus from which we infer information on the state
$\rho$ from the probability distribution $p(e|\rho)$. Positivity of $P_e$ is needed for positivity
of $p(e|\rho)$, whereas normalization is guaranteed by the completeness relation $\sum_{e\in\sE}
P_e=I$. In the present paper we will only consider the simple case of finite discrete set $\sE$.
More generally, one has an infinite probability space $\set{E}$ (generally continuous), and in this
context the set of positive operators $\{P_e \}$ becomes actually a positive operator valued measure
(POVM), but we will keep the same acronym also for the discrete case, as usual in the literature.
Every apparatus is described by a POVM, and, reversely, every POVM can be realized in principle by
an apparatus \cite{Holevo,Davies,Busch}.  Throughout this paper we will consider a quantum system
with Hilbert space $\sH$ with finite dimension $d=\dim(\sH)<+\infty$.

In the following we define the {\em data processing} $c^X_i$ for a
POVM in order to reconstruct the ensemble average $\<X\>$ of an
operator $X\in{\mathcal L}(\sH)$ ($c^X:i\mapsto c_i^X$ is the so-called {\em
  processing function}).

\subsection{Informationally complete measurements}

We recall that the space of Hilbert-Schmidt operators is isomorphic to
$\sH^{\otimes 2}$, and coincides with the space ${\mathcal L}(\sH)$ of
linear operators on $\sH$ for finite dimensional Hilbert space
$\sH\sim\Cmplx^d$.

A POVM $\bP$ is called informationally complete \cite{prug} if it
linearly spans the whole operator space ${\mathcal L}(\sH)$. We
generalize this concept to the following notion of partially
informationally complete POVM
\begin{definition} For ${\mathcal R}$ a linear operator space, we will
  call a POVM ${\mathcal R}$-informationally complete, if ${\mathcal
    R}\subseteq \Span(\bP)$.
\end{definition}
We have used the natural notation $\Span(\bP)=\Span(P_1,P_2,\ldots
P_N)\in{\mathcal L}(\sH)$.  The projection on the linear operator space ${\mathcal R}$
will be denoted by $\Pi_{\mathcal R}$.

It is clear that the knowledge of probabilities of an $\mathcal
R$-informationally complete POVM allows the calculation of ensemble
averages $\<X\>_\rho$ for all $X\in\mathcal R$ by the simple formula
\begin{equation}\label{procefun}
\<X\>_\rho=\sum_{i=1}^Nc_i^X\Tr[\rho P_i].
\end{equation}
$c_i^X$ denoting the data processing for $X$. Eq. (\ref{procefun}) has to be regarded as the
definition itself of the processing function $c^X$, in the sense that the coefficients $c_i^X$ must
satisfy Eq. (\ref{procefun}) as a constraint.  If the POVM elements are linearly independent then
the processing function $c^X:i\mapsto c_i^X$ for an operator $X$ is unique, whereas for linearly
dependent POVM elements the possible choices are infinite (notice that even thorough Eq.
(\ref{procefun}) explicitly contains the processing function, its value is independent of the
specific choice of $c_i^X$). These facts determine two questions: a) how to find a suitable
processing function $c^X$ for a given operator $X$; b) which is the processing function $c^X$
minimizing the statistical error
\begin{equation}\label{staterr}
\delta^2_\rho(X)=\sum_{i=1}^N \left|c_i^X\right|^2\Tr[\rho P_i]-\<X\>_\rho^2,
\end{equation}
where, for simplicity, we restrict to selfadjoint $X$ (notice that the actual statistical error is
obtained by dividing $\delta_\rho(X)$ by $\sqrt{N_{ex}-1}$, with $N_{ex}$ the number of
experiments). In order to answer these questions we will consider some elementary results in frame
theory.

\subsection{Elements of frame theory}

A frame in a Hilbert space $\sK$ \cite{ds,czz} (for the sake of simplicity we will consider finite
dimensional spaces) is a set of vectors $\{v_i\}_{1\leq i\leq N}\subseteq\sK$, with $N\leq\infty$
such that there exist two constants $0<a\leq b<\infty$ and
\begin{equation}
  a\N{\psi}^2_{\sK}\leq\sum_{i=1}^N|\<v_i|\psi\>|^2\leq b\N{\psi}_{\sK}^2,
\end{equation}
and one can prove that for finite dimensional systems the property of
a set $\{v_i\}$ of being a frame is equivalent to completeness, namely
for all $\psi\in\sK$ one can expand $\psi$ on the vectors $\{v_i\}$ by
suitable coefficients. On the other hand, given a set of vectors
$\{v_i\}$ on $\sK$ they are a frame iff the {\em frame operator}
\begin{equation}
  F=\sum_{i=1}^N|v_i\>\<v_i|,
\end{equation}
is bounded and invertible. In this case, defining the {\em canonical
  dual frame} $\{w_i\}$ by $F^{-1}|v_i\>=|w_i\>$ one has
\begin{equation}
  FF^{-1}=\sum_{i=1}^N|v_i\>\<w_i|=I,
\end{equation}
and clearly the coefficients $\<w_i|\psi\>$ are suitable for the
expansion of $\psi$ on the frame $\{v_i\}$, namely
\begin{equation}
  |\psi\>=\sum_{i=1}^N|v_i\>\<w_i|\psi\>.
\end{equation}
The second interesting result \cite{li} is the following
classification of all possible {\em alternate dual frames} $\{z_i\}$
such that $\sum_{i=1}^N|v_i\>\<z_i|=I$, which are given by 
\begin{equation}\label{candu}
  |z_i\>=|w_i\>+|y_i\>-\sum_{j=1}^N|y_j\>\<v_j|w_i\>,
\end{equation}
where $\{y_i\}\subseteq\sK$ is arbitrary. If we now consider the POVM $\bP$ and
$\sK\equiv{\Span(\bP)}\subseteq{\mathcal L}(\sH)$, clearly the POVM elements are a frame for
${\Span(\bP)}$, and a suitable processing function $c^X$ for an operator $X$ is provided by the
canonical dual frame. This answers the first question about finding processing functions. In the
next section we will use the classification of alternate duals in Eq.~\eqref{candu} to answer the
second question about the minimization of the statistical error.
\subsection{Optimization of the processing function}\label{opt}

The quantity we want to minimize is the statistical error in
Eq.~\eqref{staterr}. Since the processing function is involved only in
the first term, the quantity to be minimized is the following
\begin{equation}
\delta^2_\rho(X)+\<X\>_\rho^2=\sum_{i=1}^N\left|c^X_i\right|^2\Tr[\rho P_i].
\end{equation}
This quantity depends on the state $\rho$, but in a Bayesian framework
we can make it independent of $\rho$ by suitably averaging
Eq.~\eqref{staterr} over a prior ensemble ${\mathcal
  E}=\{q_j,\rho_j\}_{1\leq j\leq M}$, obtaining
\begin{equation}\label{averr}
  \delta^2_{\mathcal E}(X)=\sum_{i=1}^N\left|c^X_i\right|^2\Tr[\rho_{\mathcal E}P_i]-\overline{\<X\>^2}_{\mathcal E}=\sum_{j=1}^Mq_j \delta^2_{\rho_j}(X),
\end{equation}
where $\rho_{\mathcal E}:=\sum_{j=1}^Mq_j\rho_j$, and $\overline{\<X\>^2}_{\mathcal
  E}:=\sum_{j=1}^Mq_j\<X\>_{\rho_j}^2$.  The only term depending on the processing function is
$\sum_{i=1}^N\left|c^X_i\right|^2\Tr[\rho_{\mathcal E}P_i]$, which can be viewed as a norm for the
vector $c^X$ of coefficients in a Euclidean space $\sK$, where the metric matrix $\pi$ is diagonal
on the canonical basis and has eigenvalues $\pi_{ii}=\Tr[\rho_{\mathcal E}P_i]$. We can now define
the linear operator $\Lambda:\sK\to{\Span(\bP)}$ such that
\begin{equation}
  \Lambda c=\sum_{i=1}^Nc_iP_i,
\end{equation}
which has the following matrix elements $\Lambda_{mn,i}=(P_i)_{mn}$,
and all the generalized inverses $\Gamma:{\Span(\bP)}\to\sK$ of
$\Lambda$ satisfying $\Lambda\Gamma\Lambda=\Lambda$ are in
correspondence with alternate duals $\vec D$ by the identity
$\Gamma_{i,mn}=(D_i^*)_{mn}$. Generalizing the proof for the minimum
norm pseudoinverse in Ref. \cite{bapat} it was proved in Ref.
\cite{locvs} that the minimum norm is achieved by $\Gamma$ satisfying
\begin{equation}\label{minnorm}
  \pi\Gamma\Lambda=\Lambda^\dag\Gamma^\dag\pi,
\end{equation}
and the corresponding optimal dual was derived in Ref. \cite{optproc},
and can be expressed as follows
\begin{equation}
D_i=\Delta_i-\sum_{j=1}^N\{[(I-M)\pi(I-M)]^{\ddag}\pi\}_{ij}\Delta_j,
\end{equation}
where $\{\Delta_i\}$ is the canonical dual and the projection matrix
$M$ has matrix elements $M_{ij}=\Tr[\Delta_i P_j]$. The symbol
$Y^\ddag$ denotes the Moore-Penrose generalized inverse of $Y$, namely
the symmetric, minimum norm and least squares generalized inverse $Z$
satisfying the conditions: $ZYZ=Z$, $ZY=Y^\dag Z^\dag$,
$YZ=Z^\dag Y^\dag$. In the following we will make use of the following
compact formula for the minimum noise, which was derived in Ref.
\cite{last}
\begin{equation}\label{minerr}
  \delta^2_{\mathcal E}(X)=\<X|(\Lambda\pi^{-1}\Lambda^\dag)^{-1}|X\>-\overline{\<X\>}_{\mathcal E},
\end{equation}
where $|X\>\in\sH^{\otimes 2}$ is the vector corresponding to $X$ as follows
\begin{equation}
  |X\>:=\sum_{m,n=1}^d X_{mn}|m\>\otimes |n\>\leftrightarrow X,
\end{equation}
for fixed basis $\{|m\>\}_{1\leq m\leq d}$ in $\sH$.  
The following identities are easily verified
\begin{equation}
  \<X|Y\>=\Tr[X^\dag Y],\ A\otimes B|X\>=|AXB^T\>,\ E|X\>=|X^T\>,
\end{equation}
where $X^T$ is the transpose of $X$ in the canonical basis, and $E$ is the swap operator
$E|\phi\>\otimes|\psi\>=|\psi\>\otimes|\phi\>$.  Throughout the paper we will use the following
notation for orthogonal projectors over Hilbert-Schmidt subspaces ${\mathcal S}\subseteq{\mathcal L}(\sH)$
\begin{equation}\label{notationP}
\Pi_{{\mathcal S}}:=\text{orthogonal projector over } \Span\{|X\>,\;X\in {\mathcal S}\}.
\end{equation}

Since the POVM $\bP$ is selfadjoint, namely
$E|P_i^*\>=|P_i\>$ ($F^*= (F^\dag)^T$ denotes the complex conjugated operator), its frame operator
$F=\sum_{i=1}^N|P_i\>\<P_i|$ enjoys the following property
\begin{equation}
EF^*E=F,
\end{equation}
which is clearly shared by its inverse and by $\Pi_{\mathcal
  S}=F^{-1}F$. The canonical dual $\{\Delta_i\}$ satisfies then the
following identity
\begin{equation}
  E|\Delta_i^*\>=EF^{-1*}|P_i^*\>=F^{-1}E|P_i^*\>=F^{-1}|P_i\>=|\Delta_i\>,
\end{equation}
namely $\Delta^\dag_i=\Delta_i$. Since all alternate duals $\vec D$
satisfy
\begin{equation}
  \sum_{i=1}^N|D_i\>\<P_i|=\Pi_{\Span(\bP)}=E\Pi_{\Span(\bP)}^* E=\sum_{i=1}^N|D_i^\dag\>\<P_i|,
\end{equation}
it is clear that if $\vec D$ is an alternate dual then also
$\vec D^\dag$ is. It is easy to verify that also $1/2(D_i+D_i^\dag)$
is an alternate dual. Suppose now that the optimal dual is not
selfadjoint, then there exists a selfadjoint $X$ such that
$\Im(\Tr[D^\dag_i X])\neq0$, and the minimum statistical error for $X$
would be
\begin{equation}
\begin{split}
  \delta^2_{\mathcal E}(X)=&\sum_{i=1}^N|\Tr[D^\dag_i
  X]|^2\Tr[\rho_{\mathcal E}P_i]-\overline{\<X\>}_{\mathcal E}>\\
  &\sum_{i=1}^N\Re(\Tr[D^\dag_i
  X])^2\Tr[\rho_{\mathcal E}P_i]-\overline{\<X\>}_{\mathcal E}=\\
  &\sum_{i=1}^N(\Tr[(D^\dag_i+D_i) X]/2)^2\Tr[\rho_{\mathcal
    E}P_i]-\overline{\<X\>}_{\mathcal E}.
\end{split}
\end{equation}
This is clearly absurd, since the last line is the statistical error
given by the dual $(D_i+D_i^\dag)/2$. The canonical dual and the
optimal dual for any ensemble are then selfadjoint.\par

Writing the matrix elements of both sides in Eq.~\eqref{minnorm}, and
considering that $\Gamma\Lambda_{ij}=\Tr[D^\dag_i P_j]$, one has
$\pi_{ii}\Tr[D^\dag_i P_j]=\Tr[P_i D_j ]\pi_{jj}$. Summing both sides
over the index $i$ we obtain $\Tr[\rho_{\mathcal
  E}P_j]=\Tr[D_j]\Tr[\rho_{\mathcal E}P_j]$, and consequently
$\Tr[D_i]=1$ for all $i$ such that $\Tr[P_i\rho_{\mathcal E}]\neq 0$.

\subsection{Post-processing}

We will call {\em post-processing} of a POVM a data-processing which maps the
POVM into another POVM, namely
\begin{equation}\label{s:post-processing}
Q_j=\sum_{i=1}^N m(j|i)P_i,
\end{equation}
where $m(j|i)$ is a conditional probability, namely the corresponding matrix is Markov, \ie
$m(j|i)\geq0$ and $\sum_jm(j|i)=1$ $\forall i$. Clearly the post-processing is a special case of
data-processing array, corresponding to
\begin{equation}
c_i^{Q_j}\equiv m(j|i).
\end{equation}
Even though it can be regarded as a special case of data-processing, the post-processing is
conceptually very different, being the randomization of set-theoretical operations. Indeed, it
corresponds to a randomization of the following operations
\begin{enumerate}
\item[T1] identification of two outcomes, e.~g. $j$ and $k$ are identified
  with the same outcome $l$, corresponding to $m(n|j)=\delta_{ln}$ and $m(n|k)=\delta_{ln}$;
\item[T2] permutation $\pi$ of outcomes, corresponding to $m(\pi(j)|k)=\delta_{jk}$;
\item[T3] splitting of one outcome $l$ into two outcomes $j$ and $k$, corresponding to choosing $j$ with
  probability $m(j|l)=p$ and $k$ with probability $m(k|l)=1-p$, $0<p<1$.
\end{enumerate}
We can see that generally the cardinality of $\bQ$ is different from that of $\bP$.  Also, notice
that a data processing array $c_i^{Q_j}$ for the POVM $\bQ$ is not necessarily a Markov matrix,
since generally $c_i^{Q_j}\not\geq 0$, and also one not necessarily has normalization $\sum_j
c_i^{Q_j}=1$ $\forall i$, due to linear dependence of the POVM $\bP$, even though, there always
exists an alternate data processing that is normalized.

When two POVMs $\bP$ and $\bQ$ are connected by post-processing we will write $\bP\succ\bQ$, and say
that the POVM $\bP$ is {\em cleaner under post-processing}---{\em post-processing cleaner} in
short---than the POVM $\bQ$. The relation $\succ$ is a pseudo-ordering, since it is {\em i)}
reflexive, corresponding to $\bP\succ\bP$, $m(i|j)=\delta_{ij}$; {\em ii)} transitive, i.~e.
$\bP\succ\bQ\succ\bR$, corresponding to $R_i=\sum_jm(i|k)Q_k,\;Q_k=\sum_jm'(k|j)P_j,\Rightarrow
R_i=\sum_jm''(i|j)P_j$, $m''(i|j)=\sum_km(i|k)m'(k|j)$.
\par We can define a partial ordering and an equivalence relation in terms of the POVM post-processing as
follows.
\begin{definition}
  The POVM's $\bP$ and $\bQ$ are {\em post-processing equivalent}---in
  symbols $\bP\simeq \bQ$---iff both relations $\bP\succ\bQ$ and
  $\bQ\succ\bP$ hold.
\end{definition}
We are now in position to define {\em cleanness under post
  processing}, namely
\begin{definition}
  A POVM $\bP$ is {\em post-processing clean} if for any POVM $\bQ$
  such that $\bQ\succ\bP$, then also $\bP\succ\bQ$ holds, namely
  $\bP\simeq\bQ$.
\end{definition}
The characterization of cleanness under post-processing is very simple, and is given by the
following theorem.
\begin{theo}\cite{clean}
A POVM $\bP$   is post-processing clean iff it is rank-one.
\end{theo}
We address the reader to Ref. \cite{clean} for the proof of the Theorem.
\medskip
\par For a POVM $\bQ$ with $\bQ\not\prec\bP$ \ie which is not a post-processing of $\bP$ one can
anyway introduce another {\em smeared-out} version $\tilde\bQ$ of $\bQ$ 
\begin{equation}\label{s:smeared-out}
\tilde Q_j:=\frac{Q_j+\alpha_jI}{1+\sum_l\alpha_l},\qquad \alpha_j=\max_i\{0,-c_i^{Q_j}\}
\end{equation}
such that $\tilde\bQ\prec\bP$---\ie $\tilde\bQ$ is a post-processing of $\bP$.  The Markov matrix is
simply given by
\begin{equation}\label{s:data-processing}
m(j|i)=\frac{c_i^{Q_j}+\alpha_j}{1+\sum_l\alpha_l}.
\end{equation}

The perfect measurement of an observable corresponds to a POVM made
with the orthogonal projectors $X_j$ over its eigenspaces, and we will
write $\bX=[X_j]$ with
\begin{equation}
X_jX_i=\delta_{ij}X_i\geq 0,\quad \sum_i X_i=I.
\end{equation}
More generally, we will say that 
\begin{definition} A POVM $\bP$ describes an {\em imperfect measurement} of the observable $\bX$ if
  $\bX\succ\bP$, namely the POVM $\bP$ is a post-processing of $\bX$.
\end{definition}

In practical terms this means that the measurement is a smearing-out of the perfect observable due
to additional noise which is ascribed to the output stage of the measuring apparatus. One can see
that mathematically a POVM is a measurement of the observable $\bX$ when it commutes with the
observable.  In this way the POVM $\bP$ describing an imperfect measurement of $X$ will be simply a
function $P_i=P_i(X)$ of the operator $X$.

\par The concept of post-processing allows to introduce a general notion of
joint measurement of (generally non commuting) observables.

\begin{definition}[Joint measurement of observables]
  We say that a POVM $\bP$ achieves the joint measurement of the observables $\bX^{(1)}$,
  $\bX^{(2)}$, $\ldots$, if for every observable $\bX^{(i)}$ of the list there is a post-processing
  of $\bP$ which achieves an imperfect measurement of $\bX^{(i)}$.
\end{definition}

We stress that in our operational point of view it is irrelevant that a joint measurement is
described by a bivariate probability distribution (which could be interpreted in terms of the
alleged outcomes of the non commuting observables $A$ and $B$). The only thing that matters is the
possibility of performing jointly imperfect measurements of both $A$ and $B$, since, indeed, the
joint probability of their eigenvalues is counterfactual.

The present definition of joint measurement for different observables is sufficiently comprehensive
to include all known joint measurements, such as the joint measurement of position and
momentum\cite{ak}, and the measurement of the direction of an angular momentum, corresponding to a
POVM made with spin-coherent states \cite{perelomov}. Indeed, the usual definition of joint
measurement simply involves the marginalization of multivariate POVMs. A natural generalization of
such definition of joint measurement for non multivariate POVM's would be simply to consider the
marginalization as the identification of outcomes in T1. Our definition of joint measurements
further generalizes the notion to any post-processing, introducing also the natural transformations
T2 and T3.

We should notice that our definition (as the standard ones) of joint measurements also includes some
trivial cases, in particular: {\em a)} {\em pure guessing post-processing}, with Markov matrix with
equal columns (data processing independent of the outcome), corresponding to a smeared-out POVM
having each element proportional to the identity (clearly for such trivial smearing-out each POVM is
the joint measurement of any set of observables); {\em b)} the POVM $\bP$ achieving the joint
measurement is actually the random selection of one observable at a time, namely
$\bP=\cup_i\lambda_i\bX^{(i)}$, where we define the convex union $\bR=\lambda\bP\cup (1-\lambda)\bQ$
of two POVMs $\bP$ and $\bQ$ with cardinalities $|\bP|=N$ and $|\bQ|=M$ as follows
\begin{equation}
\begin{split}
&\bR=\lambda\bP\cup (1-\lambda)\bQ:=\\ &[\lambda P_1,\ldots,\lambda P_N,
(1-\lambda) Q_1,\ldots, (1-\lambda) Q_M],
\end{split}
\end{equation}
(more generally one can have even the random selection of imperfect measurements of noncommuting
observables).  In the following we will call the above joint measurements {\em trivial}.

\subsection{Measuring a POVM by another POVM} A special case of
processing is the one corresponding to another POVM $\vec
Q=(Q_1,Q_2,\ldots,Q_M)$ in the span $\Span(\bP)$.  Notice that, even
though one has the linearity of processing functions
$c_i^{X+Y}=c_i^X+c_i^Y$, for linearly dependent POVM $\bP$ the
processing function is not unique, whence, generally $c_i^I\neq 1$,
which implies that the processing function $c_i^{Q_j}$ for the POVM
elements $Q_j$ generally do not satisfy the normalization condition
$\sum_{j=1}^M c_i^{Q_j}=1$. In addition, generally for $X\geq 0$ not
necessarily one has $c^X_i\geq 0$.  This implies that $c_i^{Q_j}$
cannot be treated as conditional probabilities $p(j|i):=c_i^{Q_j}$.
Therefore, it is not generally true that a POVM $\bQ\in\Span(\bP)$ can
be achieved as a post-processing of $\bP$.  However, even though $\bQ$
cannot be obtained in this way, this is possible for a blurred version
of it according to the following theorem
\begin{theo}\label{th:QQ} Given a POVM $\bQ\in\Span(\bP)$, there exists a POVM $\bQ'\prec \bQ$ 
  that is a post-processing of $\bP$, or, in other words, $\bQ'\prec \bQ$ and $\bQ'\prec \bP$ .
\end{theo}
\Proof As shown at the end of Sec. \ref{opt}, the normalization requirement is satisfied at least by the
optimal processing, since $c^I_i=\Tr[D_i I]=1$ for all $i$, for the optimal dual $\bD$ of $\bP$.
For $c_i^{Q_j}\not\geq 0$, we can consider the ``blurred'' POVM $\vec
Q(\varepsilon)$ with $Q_i(\varepsilon)=(1-\varepsilon) Q_i + \varepsilon \frac{I}{M}$, which, for
sufficiently large $\varepsilon>0$ has $c_i^{Q_j(\varepsilon)}\geq 0$. The minimum value of
$\varepsilon$ is $\varepsilon_*=-\frac{M\bar c}{1-M\bar c}$, where $\bar c=\min\{0,\min_{i,j}
\{c_i^{Q_j}\}\}$.\qed

\medskip

How can we interpret the indirect measurement of $\vec Q$?  In our
approach to the theory of statistics of quantum measurements the POVM
represents a question asked by the experimenter, and the answer is the
outcome. A POVM $\vec Q$ in the space $\Span(\bP)$
associated to the POVM $\bP$ is a question that can be indirectly
asked through the POVM $\bP$, corresponding to the following rule:
for given outcome $i$ of the POVM $\bP$ pick the answer $j$ out of
the set $1,\ldots M$ randomly according to the conditional probability
$p(j|i)=c_i^{Q_j(\varepsilon_*)}$.

If we collect the statistics for the answers $j$ obtained through this
strategy, we asymptotically obtain the probabilities
\begin{equation}
\Tr[\rho Q_j(\varepsilon_*)]=(1-\varepsilon_*)\Tr[\rho Q_j]+\frac{\varepsilon_*}M.
\end{equation}
The estimated probabilities are not exactly $\Tr[\rho Q_j]$, but since
$\varepsilon_*$ is exactly known, one can retrieve $\Tr[\rho Q_j]$ by the
formula
\begin{equation}
\Tr[\rho Q_j]=\frac1{1-\varepsilon_*}\left(\Tr[\rho Q_j(\varepsilon_*)]-\frac{\varepsilon_*}M\right).
\end{equation}
The statistical error on such estimate of $\Tr[\rho Q_j]$ is now given
by $\sum_{i=1}^N\left|c_i^{Q_j}-\Tr[\rho Q_j]\right|^2 \Tr[\rho_{\mathcal E}P_i]$, and since
\begin{equation}
c_i^{Q_j}=\frac1{1-\varepsilon_*}\left(c_i^{Q_j(\varepsilon_*)}-\frac{\varepsilon_*}M\right),
\end{equation}
the statistical error in the estimate of the probabilities $\Tr[\rho
Q_j]$ is just $\frac1{(1-\varepsilon_*)^2}$ times greater than the
statistical error in the estimate of $\Tr[\rho Q_j(\varepsilon_*)]$,
and the estimated probability $\Tr[\rho Q_j]$ is unbiased.\par
Moreover, if the POVM $\vec Q$ is the spectral decomposition of an
operator $X$, then one can obtain $\<X\>$ by taking $\sum_{j=1}^M x_j
\<Q_j\>$. The minimum error in the estimate is the same that one would
obtain by estimating
$\<X(\varepsilon_*)\>=(1-\varepsilon_*)\<X\>+\Tr[X]\varepsilon_*/M$,
where $X(\varepsilon_*)=\sum_{j=1}^MQ'_j\lambda_j$, and then
calculating $\<X\>$ by taking
\begin{equation}
\<X\>=\frac1{(1-\varepsilon_*)}\left(\<X(\varepsilon_*)\>-\frac{\varepsilon_*}M\Tr[X]\right).
\end{equation}
Notice that the coefficients $c_i^{Q_j}$ can then be interpreted as
matrix elements of a linear transformation that brings eigenvalues
$x_j$ of $X$ to the processing function for X
$c_i^X=\sum_{j=1}^Mc_i^{Q_j}x_j$. If the $c_i^{Q_j}$ are evaluated
through the optimal dual, we can say that $c_i^X$ is the best estimate
of $X$ provided that the outcome $i$ has occurred in a measurement of
the POVM $\bP$, since the estimate of $\<X\>$ rising from this
strategy has the minimum statistical error.

\par From Theorem 7 and Definitions 1 and 6, it follows immediately that 
\begin{theo} Every $\mathcal{R}$-informationally complete measurement is an an unbiased joint
  measurement of all observables in
  $\Span(\mathcal{R}\cup\mathcal{R}^\dag)$, 
\end{theo}
where we denoted by $\mathcal{R}^\dag$ the linear space spanned by the adjoints of
all operators in $\mathcal{R}$. 
Moreover, one has
\begin{corollary}
Every informationally complete measurement is a nontrivial joint measurement of all observables.
\end{corollary}

\section{$AB$-informationally complete measurements}
The problem of estimating the full probability distribution of two
noncommuting observables $A$ and
$B$ can be treated by considering the space spanned by independent
powers of $A$ and $B$, which we call $AB$-{\em space}
\begin{equation}
{\mathcal S}_{AB}=\Span\{A^n,\, B^n,\, n=0,1,2,\ldots\}.
\end{equation}
The corresponding projection (in the sense of Eq. (\ref{notationP}))
will be denoted by $\Pi_{AB}$.  The POVMs allowing for simultaneous
measurement of $A$, $B$, and their independent powers 
are what we call {\em $AB$-informationally
  complete measurements}, whose space $\Span(\bP)$ contains ${\mathcal
  S}_{AB}$.

Usually in the literature, a self-adjoint operator $X=\sum_ix_iX_i$ is
associated to the observable $\bX$, and the probability distribution
$p(i|\rho)=\Tr[X_i\rho]$ is recovered by the moments of $X$ through
the set of eigenvalues $x_i\in\Reals$. 
The relation between probabilities and moments passes
through the identity
\begin{equation}
  X_h=\sum_{j=0}^{s-1}{\bf W}_{jh}X^j=\sum_{j=0}^{s-1}\sum_{k=1}^s{\bf W}_{jh}x_k^jX_k
\end{equation}
whence $\sum_{j=0}^{s-1}{\bf W}_{jh}x_k^j=\delta_{hk}$, namely ${\bf W}$ is the inverse of the
Vandermonde matrix ${\bf W}^{-1}=\{x_k^j\} $.  Linear independence of the first $s-1$ powers of $X$
(and linear dependence of any higher power), where $s$ is the cardinality of the spectrum of $X$,
follows from the fact that the minimal polynomial of $X$ $$m_X(x)=\prod_{h=1}^s(x-x_h)$$ vanishes as
$m_X(X)=0$, and it is the minimal degree polynomial vanishing at $X$, whence all powers $X^n$, $0\leq
n\leq s-1$, and only such powers, are linearly independent.

Using Theorem \ref{th:QQ} we see that there always exist two data processing of an $AB$-infocomplete
measurement giving two unbiased Abelian POVMs commuting with $A$ and $B$, respectively.  Therefore,
one has
\begin{corollary}
Every $AB$-informationally complete measurement is an unbiased joint
measurement of observables $A^n$ and $B^n$, for all integer $n$. 
\end{corollary}

A special case is that of {\em minimal} $AB$-informationally complete
POVMs, whose space $\Span(\bP)$ exactly coincides with ${\mathcal
  S}_{AB}$.  Notice that an example of $AB$-informationally complete
POVM is readily given by the union of the two orthonormal resolutions
of $A$ and $B$ with a rescaling by a factor $\frac{1}{2}$. From this
example we can conclude that the projection $\Pi_{AB}$ also enjoys the
property $E\Pi_{AB}^* E=\Pi_{AB}$. We can translate the two properties
of simultaneous measurements and $AB$-informationally complete
measurements as follows:
\begin{enumerate}
\item $\bP$ is $AB$-informationally complete iff
\begin{equation}
\Pi_{AB}\Pi_{\Span(\bP)}=\Pi_{\Span(\bP)}\Pi_{AB}=\Pi_{AB}.
\end{equation}
\item $\bP$ is {\em minimal} $AB$-informationally complete iff
\begin{equation}
  \Pi_{\Span(\bP)}\equiv\Pi_{AB}.
\end{equation}
\end{enumerate}

Notice that a joint measurement of $A$ and $B$
is generally non minimal, e.~g. it provides also estimation of correlations,
which is the case of the joint measurement of position and momentum
which minimizes the product of uncertainties \cite{ak}, or of the
covariant measurement of the angular momentum \cite{perelomov}. We
conjecture that the minimum-error POVM's belong to the set of {\em
  minimal} $AB$-informationally complete POVMs. In the next session we
will show that for $\rho_{\mathcal{E}}\in{\mathcal S}_{AB}$ the
conjecture is true for dimension $d=2$. Moreover, we have the
following \medskip

\begin{lemm}
Sufficient conditions for minimality of the optimal $AB$-informationally
    complete POVM $\vec Q$:
\begin{enumerate} 
\item the state $\rho_{\mathcal{E}}$ belongs to $\mathcal{S}_{AB}$;
\item there exists an optimal POVM $\bP$ which is $AB$-informationally complete, and such that 
the operators $Q_i$ given by $|Q_i\>=\Pi_{AB}|P_i\>$ are all positive.  
\end{enumerate}
\end{lemm}
\noindent{\bf Proof.} 
Let us consider the minimum error in Eq.~\eqref{minerr}, and recall that we are interested in
operators $X$ such that $\Pi_{AB}|X\>=|X\>$. Then
\begin{equation}
\delta_{\mathcal{E}}(X)=\<X|\Pi_{AB}(\Lambda\pi^{-1}\Lambda^\dag)^{-1}\Pi_{AB}|X\>-\overline{\<X\>}_{\mathcal E}.
\end{equation}
Since\footnote{The proof that
$\Pi_{AB}(\Lambda\pi^{-1}\Lambda^\dag)^{-1}
\Pi_{AB}\geq(\Pi_{AB}\Lambda\pi^{-1}\Lambda^\dag\Pi_{AB})^{-1}$ is
the following. Consider two positive invertible operators $X$ and $Y$ such that $X\geq Y$. Then we have
$Y^{-\frac12}XY^{-\frac12}\geq I,$
and consequently
$(Y^{-\frac12}XY^{-\frac12})^{-1}=Y^{\frac12}X^{-1}Y^{\frac12}\leq I$,
and finally $X^{-1}\leq Y^{-1}$. Now one can prove \cite{zhang} that
for a general invertible positive operator $X$ and any projection $\Pi$ one has
$  \Pi X^{-1}\Pi=(\Pi X\Pi-\Pi X[(I-\Pi)X(I-\Pi)]^{-1}X\Pi)^{-1},$
where the inverse $[(I-\Pi)X(I-\Pi)]^{-1}$ is on the support of $I-\Pi$. Now, clearly 
$$  \Pi X\Pi-\Pi X[(I-\Pi)X(I-\Pi)]^{-1}X\Pi\leq \Pi X\Pi,$$
and consequently $  \Pi X^{-1}\Pi\geq (\Pi X\Pi)^{-1}.$}
$\Pi_{AB}(\Lambda\pi^{-1}\Lambda^\dag)^{-1}
\Pi_{AB}\geq(\Pi_{AB}\Lambda\pi^{-1}\Lambda^\dag\Pi_{AB})^{-1}$,
and since
$\Lambda\pi^{-1}\Lambda^\dag=\sum_{i=1}^N\frac1{\Tr[P_i\rho_{\mathcal
    E}]}|P_i\>\<P_i|$, then we have to minimize
\begin{equation}
\begin{split}
&\<X| (\Pi_{AB}\Lambda\pi^{-1}\Lambda^\dag\Pi_{AB})^{-1}|X\>\\
=&\left\< X\left|\left(\sum_{i=1}^N\frac1{\Tr[P_i\rho_{\mathcal
        E}]}|Q_i\>\<Q_i|\right)^{-1}\right|X\right\>,
\end{split}
\end{equation}
where $|Q_i\>=\Pi_{AB}|P_i\>$.  Notice that $Q_i$ is normalized, since
\begin{equation}\label{normproj}
  \sum_{i=1}^N|Q_i\>=\Pi_{AB}\sum_{i=1}^N|P_i\>=\Pi_{AB}|I\>=|I\>,
\end{equation}
but in general $Q_i$ could not be a POVM because positivity is not
preserved by the projection $\Pi_{AB}$. However, we require $Q_i\geq0$
as a condition, whence $\vec Q$ is a POVM, and the optimal processing
is then obtained via the optimal dual of $Q_i$.

\subsection{The case of qubits}
The quantum states of a qubit are conveniently represented on the
Bloch sphere as follows
\begin{equation}
\rho=\frac 12(I+{\bf n}\cdot{\boldsymbol\sigma}),
\end{equation}
where $\boldsymbol\sigma=(\sigma_x,\sigma_y,\sigma_z)$ are the three
Pauli operators, and $\bf n$ is a vector with norm $|\!|{\bf
  n}|\!|\leq 1$. Since any positive operator is proportional to a
state, any POVM can be represented as follows,
\begin{equation}\label{pqubit}
  P_i=\alpha_i I+\beta_i\sigma_x+\gamma_i\sigma_y+\delta_i\sigma_z,
\end{equation}
where $\{\alpha_i\}$, $\{\beta_i\}$, $\{\gamma_i\}$ and $\{\delta_i\}$ are positive coefficients such that
\begin{equation}\label{posit}
  \beta_i^2+\gamma_i^2+\delta_i^2\leq\alpha_i^2,\quad\alpha_i\geq 0,
\end{equation}
and the normalization is given by
\begin{equation}
  \sum_{i=1}^N\alpha_i=1,\quad\sum_{i=1}^N\beta_i=\sum_{i=1}^N\gamma_i=\sum_{i=1}^N\delta_i=0.
\end{equation}
Notice that apart from a multiplication factor and a unitary
transformation any couple of noncommuting traceless operators $A$ and
$B$ is equivalent to the following one
\begin{equation}
  \sigma_\pm(\theta)=\sigma_x\cos\theta\pm\sigma_y\sin\theta,
\end{equation}
whose commutator is $i\sigma_z\sin2\theta$. Therefore, without loss of
generality, we will restrict attention to $\sigma_\pm(\theta)$
\footnote{Notice that it is irrelevant to add a trace to $A$ and $B$,
  since this can be done by adding an operator proportional to the
  identity, e.~g.  $X'=X+kI$, and the minimum error in the estimation
  of $\<X'\>$ would be
$$  \delta^2_{\mathcal E}(X')=\sum_{i=1}^N(c_i^X+k)^2\Tr[\rho_{\mathcal E}P_i]-(\overline{\<X\>}_{\mathcal E}+k)^2=\delta^2_{\mathcal E}(X),$$
since the processing function of the identity for the optimal
processing is $c^I_i=1$.}.  
Now $\mathcal{S}_{AB}=\Span\{\sigma_x,\sigma_y,I\}$, and we consider
the case of $\rho_{\mathcal E}\in\mathcal{S}_{AB}$. Let us take a
general POVM $\bP$ such that $\Pi_{\mathcal{S}(\vec
  P)}\Pi_{\sigma_x,\sigma_y}= \Pi_{\sigma_x,\sigma_y}$.  By
definition, such POVM is $\sigma_x,\sigma_y$-informationally complete.
We can now prove that the operators $\{Q_i\}$ defined by
$|Q_i\>=\Pi_{\sigma_x,\sigma_y}|P_i\>$ make a POVM. The normalization
can be proved as in Eq.~\eqref{normproj}. On the other hand, $Q_i$ is
positive, and this can be proved considering Eq.~\eqref{pqubit}. In
fact, acting with $\Pi_{\sigma_x,\sigma_y}$ on $P_i$ one has
\begin{equation}
  \Pi_{\sigma_x,\sigma_y}|P_i\>=|Q_i\>=\alpha_i|I\>+\beta_i|\sigma_x\>+\gamma_i|\sigma_y\>.
\end{equation}
Clearly, the conditions for positivity in Eq.~\eqref{posit} are
still satisfied. We have then proved that $\{Q_i\}$ is a minimal
$\sigma_x,\sigma_y$-informationally complete POVM. Moreover, since
$\rho_{\mathcal E}\in\mathcal{S}_{AB}$,  then
\begin{equation}
  \Tr[P_i\rho_{\mathcal E}]=\<P_i|\Pi_{\sigma_x,\sigma_y}|\rho_{\mathcal E}\>=\Tr[Q_i\rho_{\mathcal
    E}], 
\end{equation}
namely $\bP$ and $\vec Q$ give the same probability distribution over the state $\rho_{\mathcal
  E}$, whence they will have the same expectations when averaging over the ensemble ${\mathcal
  E}$. Therefore, we are in the conditions of Lemma 1, whence for optimal $\bP$ the constructed
$\vec Q$ is optimal and minimal.

From now on, we will consider POVMs $\bP$ such that
$\Pi_{\sigma_x,\sigma_y}|P_i\>=|P_i\>$. Moreover, we will restrict our
attention to ensembles with a isotropic distribution, having
$\rho_{\mathcal E}=\frac I2$. In this case
$\pi_{ii}=\Tr[P_i]/2=\alpha_i$. It is clear that we can consider rank
one POVMs, since if $P_i$ is rank 2 for some $i$, then its spectral
decomposition can be written as
\begin{equation}
  P_{i1}=\frac{\lambda_2 I}{\lambda_2-\lambda_1}-\frac{P_i}{\lambda_2-\lambda_1},\ P_{i2}=\frac{\lambda_1 I}{\lambda_1-\lambda_2}-\frac{P_i}{\lambda_1-\lambda_2},\label{onn}
\end{equation}
where $\lambda_j$ are the two eigenvalues of $P_i$. The spectral
projections belong then to the space $\sigma_x,\sigma_y$, being linear
combinations of $I$ and $P_i$. Consequently, any
$\sigma_x,\sigma_y$-informationally complete POVM can be simulated by
a rank one $\sigma_x,\sigma_y$-informationally complete, whence there
exists an optimal minimal rank-one POVM which is
$\sigma_x,\sigma_y$-informationally complete.

Rank-one minimal $\sigma_x,\sigma_y$-informationally complete POVMs
can be easily characterized by restricting the conditions in
Eq.~\eqref{posit} as follows
\begin{equation}
  \beta_i^2+\gamma_i^2=\alpha_i^2,\quad \alpha_i>0.
\end{equation}
The matrix $\Lambda\pi^{-1}\Lambda^\dag$ can be written as
\begin{equation}
  \Lambda\pi^{-1}\Lambda^\dag=\sum_{i=1}^N\frac2{\Tr[P_i]}|P_i\>\<P_i|,
\end{equation}
which is represented on the orthonormal basis
$\{\frac1{\sqrt2}|I\>,\frac1{\sqrt2}|\sigma_x\>,\frac1{\sqrt2}|\sigma_y\>\}$
in the block-diagonal form
\begin{equation}
\Lambda\pi^{-1}\Lambda^\dag=
\begin{pmatrix}
2&0\\
0&K
\end{pmatrix},
\end{equation}
with $K$ being the $2\times2$ matrix
\begin{equation}
K=2
\begin{pmatrix}
  \sum_{i=1}^N\frac{\beta_i^2}{\alpha_i}&\sum_{i=1}^N\frac{\beta_i\gamma_i}{\alpha_i}\\
  \sum_{i=1}^N\frac{\beta_i\gamma_i}{\alpha_i}&\sum_{i=1}^N\frac{\gamma_i^2}{\alpha_i}
\end{pmatrix}.
\end{equation}
The inverse can be easily calculated, and is equal to
\begin{equation}
(\Lambda\pi^{-1}\Lambda^\dag)^{-1}=
\begin{pmatrix}
  \frac 12&0&0\\
  0&\frac2{D}\sum_{i=1}^N\frac{\gamma_i^2}{\alpha_i}&-\frac2{D}\sum_{i=1}^N\frac{\beta_i\gamma_i}{\alpha_i}\\
  0&-\frac2{D}\sum_{i=1}^N\frac{\beta_i\gamma_i}{\alpha_i}&\frac2{D}\sum_{i=1}^N\frac{\beta_i^2}{\alpha_i}
\end{pmatrix},
\end{equation}
where $D=\det(K)$. 

Using this expression we can evaluate the error for $\sigma_\pm(\theta)$
\begin{equation}
\begin{split}
  \delta^2(\sigma_\pm(\theta))=&\cos^2\theta
  \left(\frac\Gamma{D}-\overline{\<\sigma_x\>^2}\right)+\sin^2\theta
  \left(\frac B{D}-\overline{\<\sigma_y\>^2}\right)\\
  &\mp2\sin\theta\cos\theta\left(\frac\Delta{D}+\overline{\<\sigma_x\>\<\sigma_y\>}\right),
\end{split}
\end{equation}
where we defined
$\Gamma:=2\sum_{i=1}^N\frac{\gamma_i^2}{\alpha_i}$,
$B:=2\sum_{i=1}^N\frac{\beta_i^2}{\alpha_i}$ and $\Delta:=-2\sum_{i=1}^N\frac{\beta_i\gamma_i}{\alpha_i}$, and consequently $D=B\Gamma-\Delta^2$.

The total error $\delta^2_{\mathcal E}(\theta):=\delta^2_{\mathcal
  E}(\sigma_+(\theta))+\delta^2_{\mathcal E}(\sigma_-(\theta))$ is
given by
\begin{equation}
  \delta^2_{\mathcal E}(\theta)=2\left[\cos^2\theta
  \left(\frac\Gamma D-\overline{\<\sigma_x\>^2}\right)+\sin^2\theta
  \left(\frac B D-\overline{\<\sigma_y\>^2}\right)\right],
\end{equation}
and we can prove by the following argument that the optimal POVM is
such that $\Delta=0$. Indeed, consider a POVM $\bP$ with given
coefficients $\alpha_i,\beta_i,\gamma_i$ corresponding to given values
for $B,\Gamma,\Delta$. Now consider the POVM $\bP'$ with the same
coefficients $\alpha'_i=\alpha_i$ and $\beta'_i=\beta_i$ as $\bP$
and with $\gamma'_i=-\gamma_i$, corresponding to $B'=B$,
$\Gamma'=\Gamma'$ and $\Delta'=-\Delta$. If we take now the POVM $\vec
P'':=\frac12(P_1,\ldots,P_N,P'_1,\ldots,P'_N)$, then the corresponding
values can be readily calculated to be $B''=B=B'$,
$\Gamma''=\Gamma=\Gamma'$ and $\Delta''=0$. Correspondingly, the
expression for the determinant $D''$ becomes $D''=B\Gamma\geq
D=B\Gamma-\Delta^2$. Since the POVM $\bP''$ can be constructed from
any POVM $\bP$, then clearly the optimal POVM minimizing the total
noise $\delta^2_{\mathcal E}(\theta)$ is such that $\Delta=0$.

Then, we have that $(\Lambda\pi^{-1}\Lambda^\dag)^{-1}$ becomes
diagonal
\begin{equation}
(\Lambda\pi^{-1}\Lambda^\dag)^{-1}=
\begin{pmatrix}
\frac12&0&0\\
0&\frac1{B}&0\\
0&0&\frac1{\Gamma}
\end{pmatrix}.
\end{equation}
For rank-one POVMs, notice that
$\sum_{i=1}^N\frac{\beta^2_i+\gamma^2_i}{\alpha_i}=1$, namely,
$B+\Gamma=2$, and the total error is given by 
\begin{equation}\label{totalerr}
  \delta^2_{\mathcal E}(\sigma_+(\theta))+\delta^2_{\mathcal E}(\sigma_-(\theta))=4\left(\frac{\cos^2\theta}{B}+\frac{1-\cos^2\theta}{2-B}-\frac{\kappa}{2}\right),
\end{equation}
with $\kappa=\frac{1}{2}(\overline{\<\sigma_+(\theta)\>^2}_{\mathcal
  E}+\overline{\<\sigma_-(\theta)\>^2}_{\mathcal E})$. The 
minimum of Eq.~\eqref{totalerr} as a function of $B$
can be easily obtained,  leading to the following bound for the total error
\begin{equation}
  \delta^2_{\mathcal E}(\sigma_+(\theta))+\delta^2_{\mathcal E}(\sigma_-(\theta))\geq2(1+\sin2\theta-\kappa).
\end{equation}
We will now provide two POVMs that achieve the bound. The first one
has the following three elements 
\begin{equation}
\begin{split}
  &P_1=p(I+\sigma_x),\\
  &P_{2\pm}=\frac{1-p}2 I-\frac p2\sigma_x\pm\frac{\sqrt{1-2p}}2\sigma_y,
\end{split}
\end{equation}
with $p=\frac{\cos\theta}{2\cos\theta+\sin\theta}$, 
and the second one has four elements
\begin{equation}
\begin{split}
  &P_{1\pm}=\frac p2(I\pm\sigma_x),\\
  &P_{2\pm}=\frac{1-p}2(I\pm\sigma_y),
\end{split}
\end{equation}
with $p=\frac{\cos\theta}{\cos\theta+\sin\theta}$.  For equal
uncertainties,  the minimum product of the r.m.s. errors is given by
\begin{equation}
\sqrt{\delta_{\mathcal{E}}^2(\sigma_+(\theta))}
\sqrt{\delta_{\mathcal{E}}^2(\sigma_-(\theta))}=
\delta_{\mathcal{E}}^2(\sigma_\pm(\theta))=1+\sin2\theta-\kappa.\label{qui}
\end{equation}
We recall that the results of the qubit case from Eq. (\ref{onn}) to
Eq. (\ref{qui}) are obtained under the assumptions of
isotropic ensemble $\rho_{\mathcal E}=\frac I2$. 
In fact, we want to stress that even in the qubit case, whenever $\rho_\mathcal
E$ corresponding to the prior ensemble is not fully lying in the space
$\sigma_x,\sigma_y$, it is not proved that the optimal POVM is
$\sigma_x,\sigma_y$-informationally complete.

\section{Conclusions}

In this paper we have introduced the concept of $AB$-informationally complete measurements, within
the context of Quantum indirect estimation theory. Compared with a customary infocomplete
measurements, the $AB$-infocomplete one in principle allows a less noisy joint estimation of all the
moments of two noncompatible observables $A$ and $B$. The concept of $AB$ can be also easily
extended to more than two observable, but we have not analyzed such generalization. We solved the
case of qubits, showing that a $\sigma_x\sigma_y$-infocomplete measurement is less noisy than any
infocomplete one.  The relation between the concept of $AB$-infocompleteness and the notion of joint
measurement of observables $A$ and $B$ has also been discussed. The relation between minimality and
optimality of $AB$-infocomplete measurements remains an open problem.

\end{document}